# OPEN LEARNING ANALYTICS: A SYSTEMATIC LITERATURE REVIEW AND FUTURE PERSPECTIVES

Arham Muslim, Mohamed Amine Chatti·, Mouadh Guesmi (Social Computing, Department of Computer Science and Applied Cognitive Science, University of Duisburg-Essen, Germany)

arham.muslim@uni-due.de, mohamed.chatti@uni-due.de, mouadh.guesmi@stud.uni-due.de

Abstract: Open Learning Analytics (OLA) is an emerging research area that aims at improving learning efficiency and effectiveness in lifelong learning environments. OLA employs multiple methods to draw value from a wide range of educational data coming from various learning environments and contexts in order to gain insight into the learning processes of different stakeholders. As the research field is still relatively young, only a few technical platforms are available and a common understanding of requirements is lacking. This paper provides a systematic literature review of tools available in the learning analytics literature from 2011-2019 with an eye on their support for openness. 137 tools from nine academic databases are collected to form the base for this review. The analysis of selected tools is performed based on four dimensions, namely 'Data, Environments, Context (What?)', 'Stakeholders (Who?)', 'Objectives (Why?)', and 'Methods (How?)'. Moreover, five well-known OLA frameworks available in the community are systematically compared. The review concludes by eliciting the main requirements for an effective OLA platform and by identifying key challenges and future lines of work in this emerging field.

## 1.    INTRODUCTION

Learning Analytics (LA) is an emerging and fast-growing data science research field that focuses on the development of methods for analyzing and detecting patterns within this data, and leverages those methods to support the learning experience. The learning analytics community has matured significantly over the past few years. To further advance learning analytics, research needs to move towards providing learning analytics at scale. Scaling up LA requires a holistic view of learning analytics on distributed datasets across a variety of different environments and contexts by applying mixed-method approaches to address a wide range of participants with diverse interests, needs, and goals. A central aspect of this discussion is the concept of open learning analytics, which represents a shift towards a new learning analytics model that takes "openness" into account. A common understanding of openness in relation to LA is still lacking in the community and research on open learning analytics platforms is still in the early stages of development. Moreover, relatively little research has been conducted so far in order to provide a systematic overview of the field and to identify various challenges that lay ahead, as well as future research directions in open learning analytics.

In this paper, we focus on openness in relation to learning analytics and strive to come up with key requirements for effective open learning analytics platforms by systematically reviewing the state of the art of LA tools available in literature as well as analyzing five emerging open learning analytics frameworks. We conclude the paper by discussing key challenges and future perspectives in the field of open learning analytics.

## 2.        OPEN LEARNING ANALYTICS

Open Learning Analytics (OLA) is a relatively new research field introduced in a proposal paper by scholars from the Society for Learning Analytics Research (SoLAR) to refer to an open platform to integrate heterogeneous learning analytics techniques (Siemens et al., 2011). The concept of "openness" is still not well defined in relation to learning analytics. The vision of openness used in the SoLAR concept paper was the need for open-source software that utilizes open processes, algorithms, and technologies enabling researchers and developers to easily integrate their own tools and methods with the platform (Siemens et al., 2011).

A first OLA summit was held in March 2014 to promote networking and collaborative research and "to bring together representatives from the learning analytics and open-source software development fields as a means to explore the intersection of learning analytics and open learning, open technologies, and open research" (Siemens, 2014). From a technical perspective, the summit focused on open system architectures and how open source communities can provide new open-source learning analytics services and products. Building on the first summit, the Learning Analytics Community Exchange (LACE) project organized in December 2014 the Open Learning Analytics Network Summit Europe to develop a shared European perspective on the concept of an open learning analytics framework (Cooper, 2014). In this summit, the most obvious aspect of open in the context of learning analytics is the reuse of code and predictive models (Sclater, 2014). In general, the main message coming out of these summits was an emphasis on "the need for open-source software, open standards, and open APIs to address the interoperability challenge in this field as well as how important tackling the ethical and privacy issues is becoming for a wide deployment of LA" (Chatti et al., 2017, p. 5).

Chatti et al. (2017) discuss in detail various perspectives of openness in the literature and provide a comprehensive list of interpretations for the term in relation to LA, including open learning; open practice; open architecture, processes, modules, algorithms, tools, techniques, and methods; open access; open participation; open standards; open research; open science; open datasets; open learner modeling; and open assessment. The authors point out that 'open' should be interpreted in relation to these conceptualizations of openness and refer to OLA as an ongoing analytics process that stresses diversity at the What-Who-Why-How dimensions. It leverages data coming from diverse sources (What?) to address various objectives (Why?) of multiple stakeholders (Who?) by applying different analytics methods (How?). We use this view of openness as a base for the discussion in the remainder of this paper.

## 3.        OPENNESS IN CURRENT LA TOOLS

A systematic analysis of the literature was conducted based on the template analysis qualitative research methodology (King, 2012) to assess and catalog the current state of tools available in the area of LA with a focus on their support for openness. The analysis was conducted in two main steps, as described below.

## 3.1    Methodology

In the first step, publications were gathered from nine online popular publication databases in the area of educational technology, namely ACM Digital Library, Education Resources Information Center (ERIC), IEEE Explore, JSTOR, Science Direct, Springer Link, Wiley InterScience, ALT Open Access Repository, and Google Scholar. The following search criteria or its equivalent was executed on each database to retrieve the publications for review:

*"Learning Analytics" AND abstract:(tool\* OR platform\* OR solution\* OR framework\* OR application\*) AND year:(>=2011)*

The specified criteria resulted in 2464 unique publications (12th of September, 2019), which were analyzed in two iterations. In the first iteration, the abstract of each publication was analyzed based on the following criteria and extraneous papers were excluded.

- Publications not available in English were excluded
- Posters, abstracts, and workshop papers were excluded unless there was a clear indication in their abstract that they provide details about an LA tool
- Publications providing empirical studies, visions, or subjective views of the authors were excluded

After the first iteration, the total number of publications was reduced to 378. Afterwards, in the second iteration, each paper was carefully read and analyzed. Moreover, the following additional criteria were applied:

- Publications discussing only abstract conceptual details of tools without providing any concrete plan for the realization were excluded
- Publications not providing details on data collection and storage, analysis, and visualization steps were excluded
- Only one publication was considered related to each tool

After the second iteration, 137 tools were identified for the final analysis. Figure 1 shows the number of reviewed tools per year.

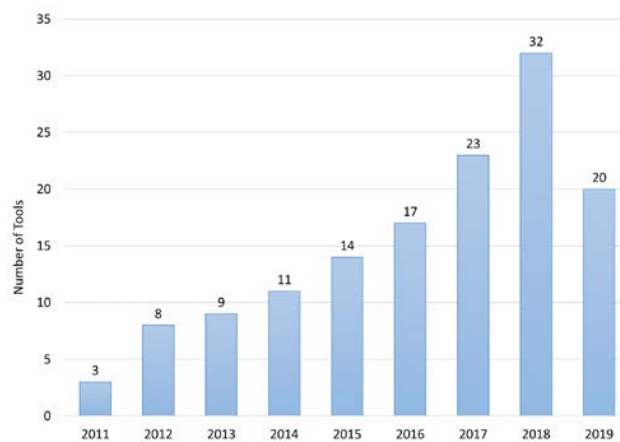

Figure 1. Number of reviewed tools per year

## 3.2        Template Analysis

In the second step, the template analysis approach was applied to thoroughly analyze and code the selected tools. The What-Who-Why-How dimensions were used as a base to formulate the template codes for the analysis of this review.

### 3.2.1        Data, Environments, Context (What?)

Data, the driving force of any LA tool, is the first template code for this analysis. In order to fully understand the different aspects of this code, it is further divided into three main sub-codes, namely data environments, data types, and data models.

*Data Environments:*
The advances in technology-enhanced learning, as well as the availability of extensive educational media, has promised the emergence of the vast amount of educational datasets from a wide variety of sources (Siemens and Baker, 2012). In order to fully cover the different types of data sources, seven classifications proposed by Chatti et al. (2012) were refined based on the discussion on LA by Siemens (2013) and Ferguson (2012). This resulted in seven data environments categories used in this review, namely 'Formal environments', 'Informal environments', 'Mixed environments', 'Ubiquitous devices', 'Student information systems', 'Social media', and 'Others'. Figure 2 shows the overview and time evolution of the coding performed on the selected tools based on the specified categories.

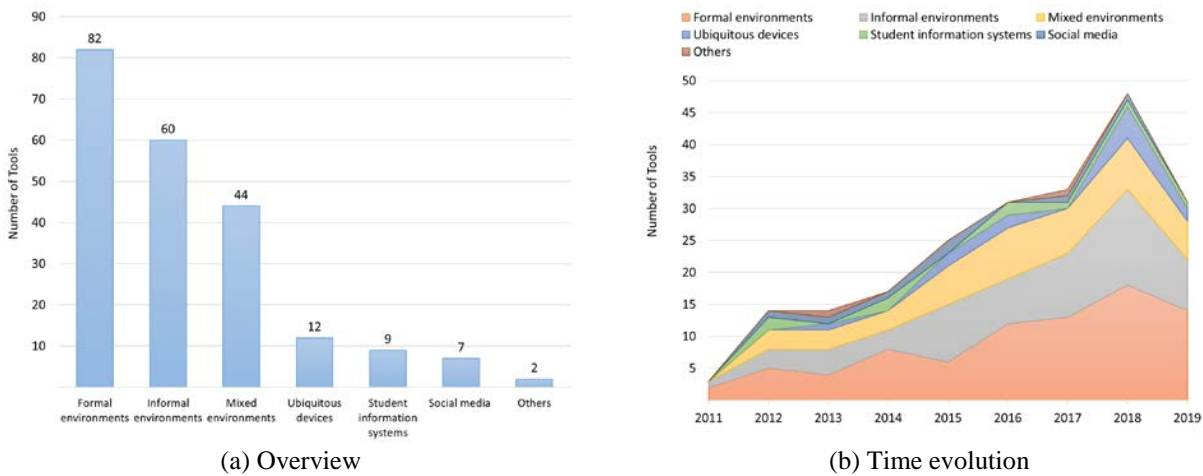

(a) Overview                                                (b) Time evolution
Figure 2. Data environments types used by the reviewed LA tools

The 'Formal environments' category represents the data sources provided by centralized learning environments of educational institutions, such as Learning Management Systems (LMS) and Virtual Learning Environments (VLE). It includes both the well-known open-source and commercial learning environments (e.g. Moodle, Blackboard, edX) as well as the custom developed ones. It is the highest-ranking category, which is used as data sources by 82 tools analyzed for this review. This is because educational institutions are the largest producers of the data sets used for LA (Ferguson, 2012). Data from 'Informal environments' was used by 60 different tools, which contains data coming from open and networked learning environments, such as Personal Learning Environments (PLE) and Massive Open

Online Courses (MOOC). This validates the statement of Chatti et al. (2012) "future LA applications will increasingly capture and mine data collected in PLEs, as a result of a shift in focus in the last few years from centralized learning systems to open learning environments" (p. 14). The increased popularity of open learning environments, e.g. MOOCs as well as the introduction of the field of OLA, has encouraged researchers to provide analytics by consolidating data from multiple learning environments. This is represented by the 'Mixed environments' category, which is used by 44 reviewed tools. However, the majority of the tools only provide support for **aggregating and integrating** data coming from a few sources together with the formal learning environment data. For instance, the tool by Conde et al. (2019) uses WhatsApp messages of students together with their activities in Moodle to identifying teamwork competencies, Edx2bigquery uses YouTube and geolocation data to provide better insights into their formal edX data (Lopez et al., 2017), and AEEA incorporates competence data based on European Qualifications Framework (EQF) in their formal dataset to provide in-depth competency analysis to educators (Florian-Gaviria et al., 2013). The 'Ubiquitous devices' is an emerging category which utilizes sensory data coming from ubiquitous handheld and wearable devices, e.g. RAP uses posture, gaze, and voice data of a student during presentation to provide automatic feedback (Ochoa et al., 2018) and SkyApp provides automated assessment to handwritten text of students based on fingers or stylus on a touchscreen of a tablet (Hui et al., 2016). Data from 'Student information systems' and 'Social media' is used by 9 and 7 tools respectively. The low values for these categories can be due to the increased concern by different stakeholders about the lack of **transparency** in the available tools and **privacy** issues. Thus, making it difficult for LA tool designers to capture data beyond LMS without explicit consent from the users (Kitto et al., 2015).

*Data Types:*

Different types of data can be collected from learning environments. Based on literature analysis, seven different data types used by LA tools are identified, namely 'Activity data', 'Assessment data', 'User profile data', 'Multimodal data', 'Social media data', 'Campus data', and 'Other dataset'.

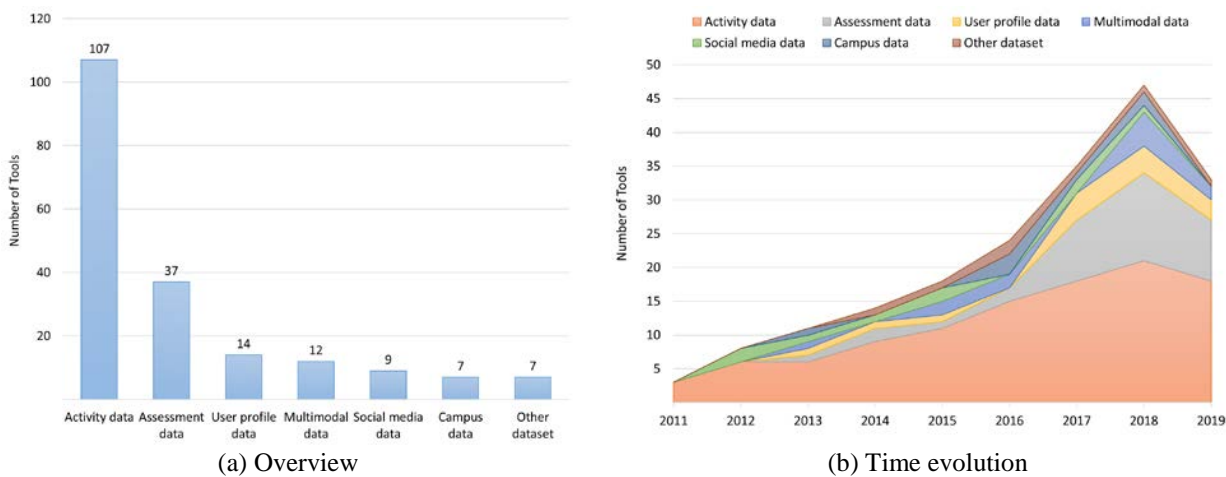

(a) Overview        (b) Time evolution

Figure 3. Data types used by the reviewed LA tools

The main data type used by almost 107 reviewed LA tools is 'Activity data', which represents the data of learning activities performed by different stakeholders available through default logging mechanism in a learning environment. It is evident from Figure 3 that analysis based on log data has been one of the focus of LA research and this trend will keep on growing with the improvements in the underlying technologies that support these environments. 'Assessment data' has emerged as a big category starting 2017 (see Figure 3b) that is used by 37 analyzed tools. This category provides more detailed data about the assessment activities performed by learners and educators that is normally not available in the default activity log data from learning environments. This includes information about obtained marks, number of solved questions, submitted solutions, textual feedback, evaluation rubrics, and sample solutions (Jayaprakash et al., 2014; Murata and Kakeshita, 2016; Syed et al., 2019). LA tools designers have also shown interested in the 'User profile data' starting 2017, which is used by 14 different tools. This category focuses on the information related to the learner's profile, including learning skills, interests, behavior tracks, demographic and aptitude data, learning and reading styles, cognitive traits, competencies, and ePortfolios(Broos et al., 2017; Dascalu et al., 2015; van der Schaaf, 2019). The 'Multimodal data' is another category that has recently gained attention. 12 analyzed tools provide analytics based on multimodal data, including audio (Griol and Callejas, 2018; Mota et al., 2018), video (Dabisias et al., 2015; Ogata and Mouri, 2015), writing (Hui et al., 2016), geo-location (Fulantelli et al., 2013), and biometric (Di Mitri et al., 2019; Tamura et al., 2019). The main reason for this increase in the usage of 'Assessment data', 'User profile data', and 'Multimodal data' is to go beyond statistics-based LA and provide more effective LA to the users by correlating various data types. For instance, using assessment data together with reading styles and cognitive traits to build learner model and support students in overcoming their reading difficulties (Mejia et al., 2017), utilizing learner's activities data together with their competencies to provide overview of competency level to educators (Vargas et al., 2019) or support learners in self-regulated learning through personalized feedback (van der Schaaf, 2019), and predicting students at risk based on their learning activities and assessment outcomes (Cobos and Olmos, 2019; Essa and Ayad, 2012; Govindarajan et al., 2015; Jayaprakash et al., 2014).

*Data Models:*
Collecting and storing data is necessary to perform any kind of analytics. The majority (110 tools) of the tools in this analysis implemented their own custom data model to manage raw data. However, **specifications and standards** available in the LA literature are also finding their way into the LA solutions to model the collected data as well as to address **integration**, **aggregation**, and **interoperability** issues. Experience API (xAPI) is at the top of the list, which is used by 22 analyzed tools to aggregate and integrate data from different sources (Alonso-Fernandez et al., 2017; Bibiloni et al., 2018; Brouwer et al., 2016; Mangaroska et al., 2019). Activity Streams (Göhnert et al., 2014; Vozniuk et al., 2013), IMS Caliper (Guenaga et al., 2015; Syed et al., 2019), and Learning Context Data Model (LCDM) (Muslim et al., 2016, 2018) are adopted by two tools each.

### 3.2.2    Stakeholders (Who?)

This template code provides an analysis of the different stakeholders who are the focus of the reviewed LA tools. This template code is divided into five main types, namely 'Educators', 'Learners', 'Administrators', 'Researchers', and 'Developers'.The overview and time evolution of the different stakeholders focused by the reviewed tools are illustrated in Figure 4.

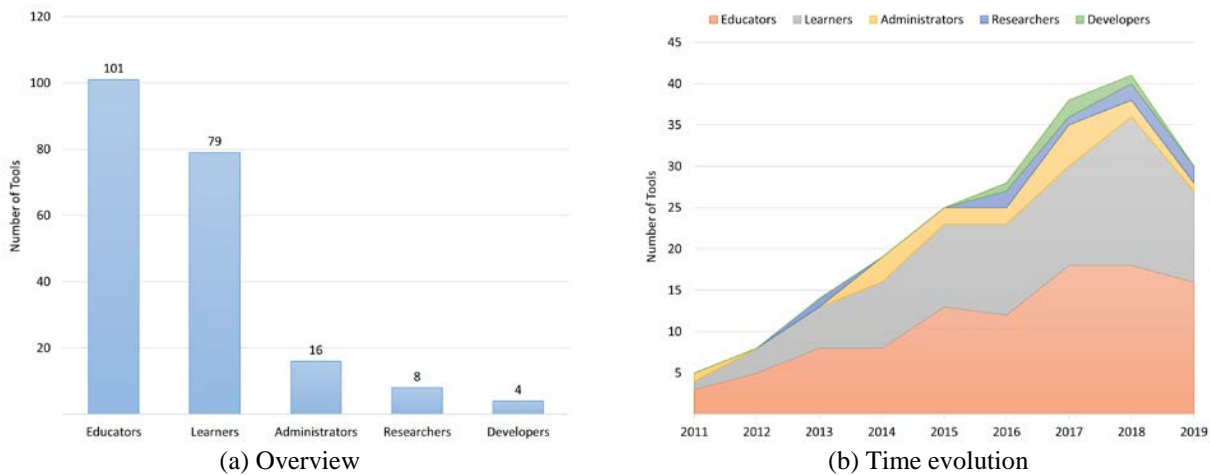

(a) Overview                    (b) Time evolution

Figure 4. Different stakeholders focused by the reviewed LA tools

The main target stakeholders of the reviewed LA solutions are 'Educators' (101 tools) and 'Learners' (79 tools). As evident from Figure 4b, the focus of LA tools is shifting towards learners with the passage of time and this trend is expected to further grow in the future. The main reason for this shift is the emergence of open and networked learning environments that allow learners to self-regulate and customize their learning experience. Institution 'Administrators' are the focus of 16 reviewed tools. Most of these tools only provided basic statistical information for monitoring activities of learners and educators. However, few tools provided more advanced analytics by supporting administrators in improving course structure (Cacatian et al., 2015; Graf et al., 2011), optimizing and load balancing open educational lab equipments (Rohloff et al., 2019), identifying students at risk (Jayakody and Perera, 2016), and enabling them to custom develop visualization for their LMS (Kapros and Peirce, 2014). To note, a considerable number of tools (56 tools) focus on multiple stakeholders at the same time, e.g. educators and learners. The majority of the surveyed tools view the stakeholders as passive users. **User involvement** is in most of the cases limited to the interaction with the provided visualizations. Only few tools actively involve users in the LA process. For example, Pardo et al. (2018) allow educators to create a basic set of rules to personalize the actions for each student and Muslim et al. (2016, 2018) enable different stakeholders to self-define custom indicators.

### 3.2.3    Objectives (Why?)

The objectives of LA have been classified into seven main categories, namely 'Monitoring and Analysis', 'Reflection and Awareness', 'Assessment and Feedback','Prediction and Intervention','Personalization and Recommendation', 'Adaptation', and 'Flexible'. Figure 5 shows the overview and time evolution of the coding performed on the selected tools.

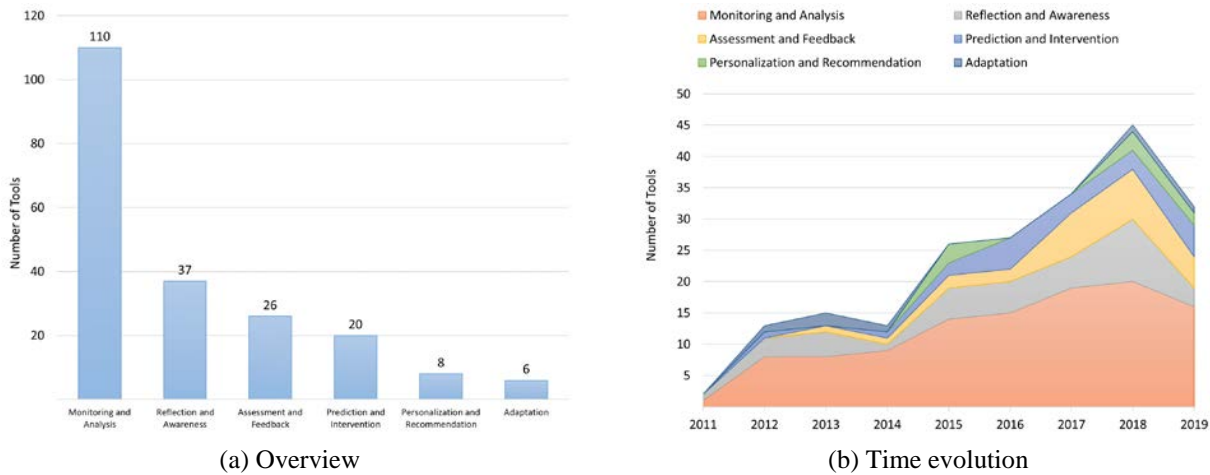

(a) Overview                                                    (b) Time evolution

Figure 5. LA objective types addressed by the reviewed LA tools

The main LA objective focused by almost 110 reviewed tools is 'Monitoring and Analysis'. The reason for this high attention is that it is relatively easy to provide basic statistical indicators for the purpose of monitoring different aspects of learners' and educators' activities. The next closely related objective is 'Reflection and Awareness', which is addressed by 37tools, which provided analytics to learners in order to improve their learning based on the overview of performed learning activities (Nitu et al., 2018; Glahn, 2013), engagement with LMS (Aljohani et al., 2019), comparison with peers in terms of performed activities (Brouwer et al., 2016), interaction and participation in blogs (Michailidis et al., 2017), and history of performed quizzes (Ogata and Mouri, 2015). Additionally, some tools focused on 'Reflection and Awareness' for educators by providing an overview of their interactions with community to increase awareness and improve learning design (Michos and Hernández-Leo, 2018) and reflecting on their competences for continuous professional development (Song et al., 2011). The 'Assessment and Feedback' objective is focused by 25 reviewed tools. A wide range of assessment types are addressed in the literature, including formal text-based assessment (Gañán et al., 2017; Kuosa et al., 2014), automatic inspection of e-portfolios (Müller et al., 2017; van der Schaaf, 2019), competency assessment (Florian-Gaviria et al., 2013; Guenaga et al., 2015), and assessing comprehension in individual and collaborative learning scenarios (Dascalu et al., 2015). The 'Prediction and Intervention' goal is focused by 20 reviewed tools, which mainly focus on predicting students' course completion time, grades, performance, drop-out rate, and retention rates (Charleer et al., 2018; Jayaprakash et al., 2014). Almost half of the reviewed tools (70 tools) address more than one objective, e.g. 'Monitoring and Analysis' together with 'Reflection and Awareness'. These are mostly the tools which are focusing on educators and learners as stakeholders. Overall, it can be seen in Figure 5b that the 'Reflection and Awareness' and 'Assessment and Feedback' objectives have gained attention starting 2017, which is due to the increase in learner-focused LA tools. In general, the surveyed tools support predefined objectives and there is a lack of **flexible** mechanisms to address new objectives of different stakeholders.

### 3.2.4    Methods (How?)

The fourth template code focuses on the techniques and methods used by the reviewed LA tools to analyze and visualize the collected data. This code is further divided into two main clusters, namely analysis types and visualization settings.

*Analysis Types:*

This cluster addresses the different types of analytics methods used by the reviewed publication to analyze collected data. It is dividing into seven main categories, namely 'Basic Data Visualization', 'Statistics', 'Classification', 'Social Network Analysis', 'Clustering', 'Others', and 'Extensible Collection'.

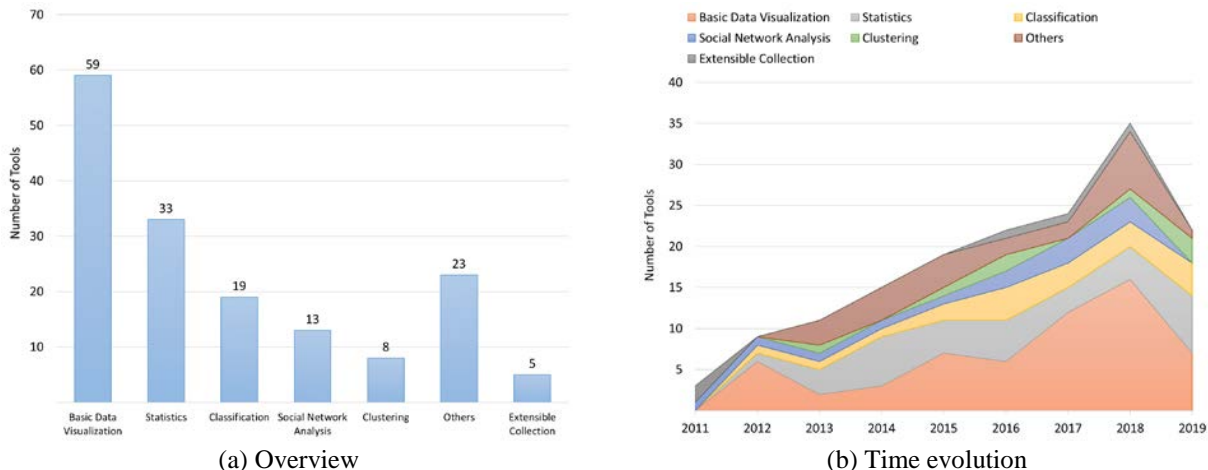

(a) Overview                                 (b) Time evolution

Figure 6. Types of analysis techniques used by the reviewed LA tools

The 'Basic Data Visualization' technique is adopted by 59 reviewed tools, as illustrated in Figure 6a. It is commonly used for the 'Monitoring and Analysis' LA objective when the objective is to perform either simple (e.g. sum, average, count) or no analysis on the raw data before visualizing it. 'Statistics' represents the second-largest category used by 33 different tools. The majority of these tools combine statistical methods with other advanced analysis techniques to provide the required LA (Bakharia et al., 2016; Carchiolo et al., 2016; Vaclavek et al., 2018). Different types of 'Classification' techniques are used by 19 reviewed tools, which focus on predicting students' performance and success based on student activity data (de Quincey et al., 2019; Essa and Ayad, 2012), identifying less-able students who are at the risk of failing (Govindarajan et al., 2015; Jayakody and Perera, 2016; Jayaprakash et al., 2014), predicting dynamic learning paths for learners (Flanagan and Ogata, 2019; Govindarajan et al., 2016), and classifying learners' behaviors to improve pedagogical practices (Hui et al., 2016). Besides the specified categories of analysis, wide variety of 'Others' analysis techniques were employed by the reviewed tools, including Natural Language Processing (NLP) (Lewkow et al., 2016; Tarmazdi et al., 2015), text mining (Dascalu et al., 2015; Kuosa et al., 2014), linear regression (Mwalumbwe and Mtebe, 2017; Nitu et al., 2018), association rule mining (Fulantelli et al., 2013), graph embedding (Giabbanelli et al., 2019), and speech recognition (Griol and Callejas, 2018). In order to improve the tool **performance** and make it **scalable** for a large amount of data, some reviewed tools also employed big data technologies and techniques, such as MapReduce (Ruipérez-Valiente et al., 2017), Hadoop Distributed File System (HDFS) (Govindarajan et al., 2015, 2016; Pan et al., 2016), Apache Spark (Lewkow et al., 2016), Apache Flink (Villanueva et al., 2018), and Google BigQuery (Lopez et al., 2017). The 'Extensible Collection' category contains 5 tools, which adopted **modular** and **reusable** architecture to allow **extensibility** of the tool with new analytics methods by supporting researchers and developers to add new Python/R based analysis script files (Bader-Natal and Lotze, 2011; Amigud et al., 2017), new Drupal-based analysis modules (Graf et al., 2011), or new analysis by following custom templates and guidelines provided by the system (Muslim et al., 2018; Lewkow et al., 2016).

*Visualization Settings:*

This cluster provides information on how the analyzed data is presented to the users, which is important for the **usability** of LA tools. This cluster is classified into six categories, namely 'Static Dashboard', 'Interactive Dashboard', 'Native Application', 'Embedded', 'Flexible Placement', and 'Others'. The overview and time evolution of the visualization setting provided by the reviewed tools is shown in Figure 7.

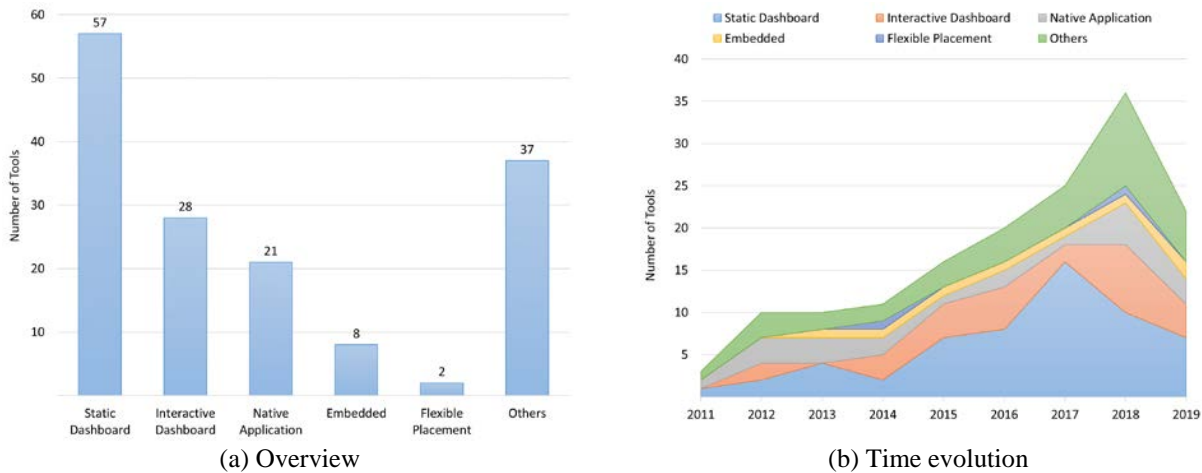

(a) Overview                                  (b) Time evolution

Figure 7. Types of visualization settings provided by the reviewed LA tools

The analysis results are shown on a web-based 'Static Dashboard' by 57 reviewed tools. These web-based dashboards are relatively easy to create due to the availability of flexible and easy to use visualization libraries like Google Charts, D3.js, C3.js, and Highcharts. Web-based 'Interactive Dashboard', provided by 28 reviewed tools, focuses on providing more exploratory LA experience to the users by providing advanced selecting, filtering, searching and sorting capabilities on the dashboard (Pesare et al., 2015; Riofrío-Luzcando et al., 2019). As compared to the dashboards, where all the analytics is provided in one place, the 'Embedded' category (supported by 8 reviewed tools) provides the analytics related to each learning activity by embedding the visualizations in its **context** (Halimi et al., 2018; Kapros and Peirce, 2014; Venant et al., 2016). Different 'Native Applications' are provided by 21 reviewed tools, which provides visualization inside LA tools developed for specific platforms, including mobile apps (Aljohani et al., 2019; Hui et al., 2016), native serious games (Malliarakis et al., 2014; Minovic and Milovanovic, 2013), and desktop applications (Giabbanelli et al., 2019; Pan et al., 2016; Tran et al., 2014). Only two tools, namely Analytics workbench (Göhnert et al., 2014) and OpenLAP (Muslim et al., 2018) support 'Flexible Placement' of analysis visualizations. These tools provide visualizations in a platform-independent HTML and JavaScript format. However, the Analytics workbench only allows adding visualizations to their own personalized widgets; whereas, OpenLAP allows embedding in any web page. Besides these common visualization settings, many 'Others' types of visualization mediums and styles have been used by different tools, e.g. textual reports (Gañán et al., 2017), tabular data (Conde et al., 2019; Mota et al., 2018), audio (Griol and Callejas, 2018), smart glasses (Holstein et al., 2018), and in-game visualizations (Perez-Colado et al., 2017). Moreover, some tools did not specify how analysis results would be provided to the users.

## 3.3    Summary

In summary, the support for openness in the analyzed tools is very limited in terms of data, stakeholders, objectives, and methods. In fact, the tools address the goals of a limited set of stakeholders, work with a specific set of data, answer a predetermined set of objectives, and rely on a predefined set of analytics methods.

## 4.        A COMPARISON OF OLA FRAMEWORKS

This section discusses the state of the art in OLA by reviewing the five most popular frameworks in the OLA community, namely SoLAR OLA, Apereo LAI, Jisc OLAA, SURFnet LAA, and OpenLAP using the main steps of the LA cycle: data collection, data storage & processing, analysis, visualization, and action (Chatti and Muslim, 2019). These frameworks are then systematically compared based on the What-Who-Why-How dimensions.

### 4.1        Society for Learning Analytics Research

The Society for Learning Analytics Research (SoLAR) proposed in 2011 the initial conceptual framework of an integrated and modularized OLA platform to integrate heterogeneous learning analytics techniques. The aim of the SoLAR OLA platform is to assist learners in tracking and improving their learning activities as well as support academics and organizations to evaluate learner activity, determine needed interventions, and improve the advancement of learning opportunities (Siemens et al., 2011).

The SoLAR OLA addresses the different steps of the LA cycle based on four core components, namely *Analytics Engine*, *Dashboard*, *Intervention Engine*, and *Learning Adaptation and Personalization Engine*. The *Analytics Engine* was planned to be the central framework to identify and process data collected from different sources based on various analysis modules. The *Dashboard* is the sensemaking component responsible for presenting visualized data to assist individuals in making decisions about teaching and learning. To support the action step in the LA cycle, the *Intervention Engine* will track learner progress and provide various automated and educator interventions using prediction models developed in the *Analytics Engine*. And, the *Learning Adaptation and Personalization Engine* will utilize the learner's profile to adjust the learning process, instructional design, and learning content to deliver customized and personalized learning materials to each individual (Siemens et al., 2011).

The focus of SoLAR was mainly directed toward creating the foundation for the theoretical framework of OLA and to provide the focal point for the educational community to engage in OLA. However, no funding was obtained to push the work forward and very little progress was made to realize the platform (Griffiths et al., 2016).

### 4.2        Apereo Learning Analytics Initiative

The Apereo foundation launched in 2014 the Apereo Learning Analytics Initiative (LAI), which proposed a diamond-shaped architecture for OLA based on five main sections; *Collection*, *Storage*, *Analysis*, *Communication*, and *Action*. The *Collection* section allows the gathering of learning activities from any xAPI and/or IMS Caliper conformant source. The collected data is stored in a Learning Record Warehouse (OpenLRW), which can support events capture with xAPI as well as IMS Caliper. The main aim of the *Analysis* section is to perform predictive analytics using the Learning Analytics Processor (LAP) component that can perform data mining, data processing, predictive model scoring, and reporting. The *Communication* section consists of a framework called 'OpenDashboard' that displays visualizations and data views called 'Cards', where each card represents a discrete visualization that shared an API and data model. In order to perform different actions, the output of the analysis can be sent to other systems, such as Student Success Plan (SSP) which includes case management, academic advising tools, early alert system, reporting and data collection tools, and integration with student information systems (Apereo, 2015; Griffiths et al., 2016; Jayaprakash, 2015).

## 4.3       Jisc Open Learning Analytics Architecture

Jisc introduced in 2015 the Open Learning Analytics Architecture (OLAA), which has much in common with Apereo LAI. In fact, Jisc OLAA is building upon Apereo's architecture with the elements collection, storage, analysis, action, and communication. The Jisc OLAA is a three-layered architecture consisting of the *Data Collection* layer, the *Data Storage and Analysis* layer, and the *Presentation and Action* layer. The *Data Collection* layer is responsible for collecting the learning activities data of students extracted from VLEs, including Moodle and Blackboard. It collects student's information from additional sources, including student record system and any self-declared data via the student app. The *Data Storage and Analysis* layer deals with storing the collected data in an xAPI compliant learning records warehouse, namely Learning Locker. The stored data is analyzed using a learning analytics processor, a tool similar to the LAP component of the Apereo LAI. Thus, it also focuses mainly on predicting the success rate of students and providing other analytics on learning data to be used by the student intervention systems. The *Presentation and Action* layer consists of different tools to visualize and provide feedback to the users based on the analysis. It includes staff dashboards, student app, and tools to manage alerts and interventions (Sclater, 2016; Griffiths et al., 2016).

## 4.4       SURFnet Learning Analytics

SURFnet started working in 2016 on the development of their LA architecture with the aim to provide teachers and students insights into study behavior. The learning analytics architecture of SURFnet can be divided into four layers; *input* layer, *data* layer, *business* layer, and *presentation* layer. The *input* layer allows collecting learner's activity data from different sources and learning environments. The *data* layer consists of an LRS to store the xAPI-based activates data. The *business* layer aggregates, organizes, and analyzes the data from the LRS to be used in the *presentation* layer. Finally, the *presentation* layer visualizes the analyzed data on dashboards and apps (Dompseler, 2016). Since the SURFnet learning analytics architecture mainly focuses on providing insights into study behavior, it lacks the action step of the LA cycle.

## 4.5       Open Learning Analytics Platform

Chatti et al. presented in 2017 a concrete conceptual and technical architecture for an open learning analytics ecosystem (OpenLAP). It provides end-users with a flexible and dynamic mechanism to generate their personalized indicators. Furthermore, to meet the requirements of diverse users, OpenLAP adapts a modular and extensible architecture to allow easy integration of new analytics objectives, analytics methods, and visualization techniques (Chatti et al., 2017; Muslim et al., 2018, 2017).

OpenLAP supports the steps of the LA cycle as follows. The data collection and storage steps are addressed by the *Data Collection & Management* component. It is also responsible for applying privacy policies to the collected data as well as generating learner and context models. The collected data is stored after aggregation and integration based on the LCDM, which is a user-centric data model that is modular and easy to understand (Muslim et al., 2018). The analysis step of the LA cycle is carried out by *Analytics Engine* together with *Analytics Modules* and *Analytics Methods* components. The *Analytics Engine* also enables extensibility in OpenLAP by providing a web service based mechanism to manage indicators, analytics goals, analytics methods, and visualization techniques. The analyzed data is visualized by the *Visualizer* component, which generates an HTML and JavaScript based code that can be placed on any web-based application. OpenLAP is silent about the action step. The possible actions are only present indirectly through the *Analytics Modules* component.

## 4.6        Comparison

In this section, the five OLA frameworks discussed above are systematically compared based on the What-Who-Why-How dimensions. The summary of the comparison is compiled in Table 1.

Table 1: OLA frameworks comparison

| | | SoLAR OLA | Apereo LAI | Jisc OLAA | SURFnet's LA | OpenLAP |
|---|---|---|---|---|---|---|
| **What?** | **Data Model** | - | xAPI, IMS Caliper | xAPI | xAPI | LCDM |
| | **Specifications** | - | LTI, PMML | LTI, PMML | OOAPI | PMML |
| **Who?** | **Stakeholders** | Learners, Educators, Administrators, Researchers | Learners, Educators | Learners, Educators | Learners, Educators | Learners, Educators, Administrators, Researchers, ... |
| **Why?** | **Objectives** | Intervention, Adaptation, Personalization | Monitoring, Prediction, Intervention | Monitoring, Prediction, Intervention | Monitoring | Monitoring, Awareness, Prediction, Reflection, Personalization, Recommendation, ... |
| | **Indicator Generation** | Developers | Developers | Developers | Developers | Stakeholders (using intuitive UI) |
| **How?** | **Analysis Methods** | SNA, Data Mining | Statistics, Visualizations, Predictive Analytics, | Statistics, Visualizations, Predictive Analytics, | Statistics, Visualizations | Statistics, SNA, Data Mining, Visualizations, … |
| | **Visualization Setting** | Dashboards | Dashboards, Native Apps | Dashboards, Native Apps | Dashboards, Native Apps | Any web-based application |
| | **Extensibility** | Tools, Methods | Predictive Models, LTI based Visualizations | Predictive Models, LTI based Visualizations | - | Analytics Goals, Analytics Methods, Visualization Techniques |

### 4.6.1        Data Models and Specifications (What?)

All the discussed frameworks utilize one or more data model specifications to aggregate the collected data, except the SoLAR OLA, which is only a conceptual framework. Apereo LAI uses the xAPI and IMS Caliper based OpenLRW. Both, Jisc OLAA and SURFnet's LA make use of the xAPI based OpenLRS; specifically, the open-source implementation called the Learning Locker. It is evident that the xAPI is becoming a de facto standard for standardizing the data collection process in OLA. OpenLAP uses LCDM as the central data model to store learning activities data.

Besides using standard data models, the selected platforms and architecture incorporate various specifications and standards. Apereo LAI, Jisc OLAA, and OpenLAP support the open model approach of OLA by using Predictive Model Markup Language (PMML) as a standard format to use and share predictive models within and outside the system. The Learning Tools Interoperability (LTI) by IMS

Global Learning Consortium is used by both Apereo LAI and Jisc OLAA to easily share visualization generated by the systems with other LTI compliant systems. SURFnet is collaborating with higher education institutions to define an Open Education API, also known as Open Onderwijs API (OOAPI) that can support educational tools to easily and efficiently communicate and share data with each other, including exercise marks, study credits, schedules, and free workstations (Ward, 2016).

### 4.6.2      Stakeholders and Objectives (Who? and Why)

The main focus of the current implementations of Apereo LAI and Jisc OLAA is to monitor and predict learners' performance to allow self-reflection and provide alerts to educators for intervention (Jayaprakash, 2015; Sclater, 2016). In the early stages of experimentation, the SURFnet LA is only focusing on monitoring various activities of learners to provide educators with insight into their behavior (Dompseler, 2016). OpenLAP focuses on provide openness through flexible definition and dynamic generation of indicators to meet the needs of different stakeholders with diverse goals beyond monitoring, prediction, and intervention. Additionally, allowing users to self-define indicators makes it possible for OpenLAP to grow its indicators catalog over time, unlike the other architectures where only developers can programmatically implement new indicators.

### 4.6.3      Methods (How?)

In terms of analysis, the focus of Apereo LAI and Jisc OLAA is to perform predictive analytics. Apereo LAI and Jisc OLAA have a common component for the analysis called LAP, which is responsible for tracking learner progress and providing various automated and educator interventions for students at risk using various prediction models (Jayaprakash, 2015; Sclater, 2016). Since OpenLAP provides a mechanism to easily add new analytics methods, it is not restricted to predictive analytics and can apply various analysis techniques (i.e. statistics, data mining, and social network analysis) to detect patterns in educational data sets.

   The impact of algorithms, visualizations, and dashboards will not be meaningful unless they are integrated into the activities they are intended to improve (Wise et al., 2014). Thus, where and how analysis results are presented to users is an important factor in maximizing the impact of LA. Most available tools present insights as indicators on dashboards or within some native applications. Following this trend, Apereo and Jisc have developed some components such as Student App, Student Success Plan (SSP), and LTI based OpenDashboard (Jayaprakash, 2015; Sclater, 2016). Similarly, SURFnet has its own dashboard and a set of apps to present analysis results to its users. The indicators in OpenLAP are not bound to any dashboard or application, rather they are available in the form of HTML and JavaScript which can be embedded in any web-based application making it possible to seamlessly integrate LA in any context.

## 5.      OLA PLATFORM REQUIREMENTS

Now, what are the requirements for an effective OLA platform? The results of the literature review and the comparison of the different OLA frameworks revealed that openness in terms of data, stakeholders, objectives, and methods requires addressing various requirements. These requirements and their relation to the What-Who-Why-How dimensions are summarized in Table 2.

Table 2: OLA platform requirements summary

| | Attributes | Description |
|---|---|---|
| **What?** | Privacy | It is crucial to build ethics and privacy into the LA solutions right from the very beginning. |
| | Transparency | Transparency is vital to drive forward the acceptance of LA. It provides an explicit definition of means how to achieve legitimacy in the LA process. It should be applied across the complete process, without exceptions. |
| | Data Aggregation and Integration | Aggregate and integrate educational data from multiple sources to create a useful educational dataset that reflects the distributed activities of the learner. |
| | Interoperability | Address the challenge of efficiently and reliably moving data and services between different educational systems. |
| | Specifications and Standards | Adopt widely accepted specifications and standards e.g. xAPI, IMS Caliper in order to achieve interoperability of datasets and services. |
| **Who?** | User involvement | Adopt a user-in-the-loop approach that involves end-users in the LA process. |
| **Why?** | Flexibility | Address the various needs and goals of different stakeholders. |
| **How?** | Extensibility | Enable extending the platform with new data sources, analytics methods, and visualization techniques. |
| | Modularity | Make it easy to accommodate new components developed by different collaborators in order to respond to changes over time. |
| | Reusability | Follow the four R's "Reuse, Redistribute, Revise, Remix" in the conceptualization and development of OLA platforms. |
| | Context | Enable to place LA indicators in context. |
| | Usability | Usability guidelines and design patterns should be taken into account. |
| | Performance and Scalability | Performance and scalability should be taken into consideration in order to allow for an incremental extension of data volume and analytics functionality. |

# 6.     OLA CHALLENGES AND FUTURE PERSPECTIVES

OLA is a highly challenging task. It raises a series of challenges and implications for LA stakeholders and there is still a great deal of research that can be done in this area. We outline in the next sections some challenges and give insights into potential next steps for OLA research. To stress here that these perspectives are also valid in learning analytics research in general, but are more challenging in open learning analytics due to the breadth of the area and the numerous aspects to address.

## 6.1     Technical Aspects

In order to further develop the field, a range of technical challenges related to the four What-Who-Why-How dimensions need to be addressed. These include the volume, velocity, and variety of educational data, triangulating multiple data sources, data privacy, and scalability. These challenges offer starting points for further research on open learning analytics technical implementations.

## 6.2      Pedagogical Aspects

The pedagogical aspects are probably the most important yet under-explored aspects of research regarding open learning analytics. These include:

### 6.2.1      Adoption

While there has been promising research on technical frameworks for open learning analytics, research on their effects is still in its early stages. if open learning analytics frameworks are to serve the intended objective of achieving learning analytics at scale, large-scale studies are to be conducted with a focus on evaluating these frameworks regarding adoption and learning impact.

### 6.2.2      Action Support

The current open learning analytics technical frameworks discussed in Section 4 are mainly focused on the technical (data and analysis) aspects instead of how to best support learners and teachers in a learning scenario. In fact, these frameworks do not propose any concrete solution for the deployment of support actions. In the future, open learning analytics research needs to widen its focus to pay attention to closing the loop by exploring the use of data to provide effective learner support actions.

### 6.2.3      Human-Centered Open Learning Analytics

Some initiatives have built elements of human-centered open learning analytics. For instance, OpenLAP provides mechanisms to help users self-define indicators according to their needs and goals; however, more work is still needed to provide effective user involvement in the design, analysis, and evaluation of open learning analytics.

## 7.      CONCLUSION

This paper offers a systematic literature review of the state of research on open learning analytics organized along the What-Who-Why-How dimensions. The review focuses specifically on how openness is supported in current learning analytics solutions. The main aim of this review was to derive requirements for effective open learning analytics platforms. The paper further identifies challenges and directions that offer new routes for open learning analytics research in the future. Along with technical aspects, there are more crucial pedagogical ones related to demonstrating the benefits for learning and teaching through active support and user involvement. Addressing these aspects can drive the adoption of open learning analytics at scale.